\documentclass[twocolumn,showpacs,preprintnumbers,amsmath,amssymb,
 aps,
 prb,
 lengthcheck
]{revtex4}
\usepackage{amssymb}
\usepackage{graphicx}
\usepackage{dcolumn}
\usepackage{bm}
\usepackage{hyperref}
\usepackage{makeidx}

\usepackage[english]{babel}
\usepackage[latin1]{inputenc}
\usepackage{color}
\usepackage{graphicx}
\usepackage{amsmath}

\usepackage{latexsym}
\usepackage{amsthm}

\makeindex
\def\>{\right\rangle}
\def\<{\left\langle}
\def\be{\begin{equation}}
\def\ee{\end{equation}}
\def\ba{\begin{array}{l}}
\def\ea{\end{array}}

\def\beq{\begin{eqnarray}}
\def\eeq{\end{eqnarray}}

\begin{document}

\preprint{APS/123-QED}

\title{Spin current pumping in helical Luttinger liquids}

\author{D. Ferraro$^{1,2,3, 4}$, G. Dolcetto$^{1,2,3}$, R. Citro$^{5, 6}$, F. Romeo$^{5,6}$, M. Sassetti$^{1, 2}$}
 \affiliation{$^1$ Dipartimento di Fisica, Universit\` a di Genova,Via Dodecaneso 33, 16146, Genova, Italy.\\
$^2$ CNR-SPIN, Via Dodecaneso 33, 16146, Genova, Italy.\\
$^3$ INFN, Via Dodecaneso 33, 16146, Genova, Italy.\\
$^4$ CNRS - Laboratoire de Physique de l'Ecole Normale Sup\'erieure de Lyon, 46 All\'ee d'Italie, 69364 Lyon Cedex 07, France.\\
$^5$ Dipartimento di Fisica "E. R. Caianiello", Universit\` a degli Studi di Salerno, Via Ponte don Melillo, I-84084, Fisciano (Sa), Italy.\\
$^6$ CNR-SPIN, UO Salerno, Via Ponte don Melillo, I-84084, Fisciano (Sa), Italy.}

\date{\today}

\begin{abstract}
\noindent
We study parametric quantum pumping in a two-dimensional topological insulator bar in the presence of electron interactions described by an helical Luttinger liquid. The pumping current is generated by two point contacts whose tunneling amplitudes are modulated in time. The helical nature of the edge states of the system ensures the generation of a pumped spin current that is determined by interference effects related to spin-flipping or spin-preserving tunneling at the quantum point contacts and which can be controlled by all electrical means.
We show that the period of oscillation and the position of the zeros of the spin current depend on the strength of the electron interactions, giving the opportunity to directly extract information about them when measured.
\end{abstract}

\maketitle
\section{Introduction}
In recent years the edge states of two dimensional topological insulators (2D TIs), showing the quantum spin Hall (QSH) effect, have been subject of attention from the condensed matter community.\cite{Konig08, Qi11} These states have been theoretically predicted in graphene layers with spin-orbit interaction \cite{Kane05a, Kane05b} and strained semiconductors \cite{Bernevig06a} and experimentally observed in Mercury-Telluride quantum wells. \cite{Bernevig06b, Konig07, Roth09} Their most relevant feature is the helicity \cite{Wu06}, namely the fact that opposite spins propagate in opposite directions along the same edge. As far as time reversal symmetry is preserved, the backscattering between these different spin channels is indeed suppressed. \cite{Hasan10} \\
Till now various proposals have been made in order to extract information about the helical nature of the edge states and their stability by means of transport measurements, including the investigation of the Kondo effect \cite{Wu06, Maciejko09, Tanaka11, Eriksson12}, the analysis of peculiar geometries like the corner junction \cite{Hou09} and the study of possible backscattering processes involving only one edge. \cite{Wu06} In the simple quantum point contact  (QPC) geometry \cite{Teo09}  signatures of helicity and electron-electron interactions can be found in the power-law behavior of the spin current as a function of the source-drain voltage in a two terminal configuration. \cite{Strom09, Dolcetto12}

Recently, also interferometric setups, similar
to the ones that have already provided hints about the existence of fractional charge and
fractional statistics in the framework of the quantum Hall (QH) effect\cite{Chamon97}, have been proposed. \cite{Dolcini11,Citro11,Virtanen11,Romeo12} In the context of QSH effect, the existence of two spin direction, with a defined helicity makes the interferometric properties of the four terminal setup, shown in Fig.\ref{fig1}, much richer. The fact that edge states with opposite spins, and belonging to different boundaries, are constrained to be close at the QPCs, as shown in Fig. \ref{fig1}, activates a local spin-orbit coupling that can be modeled phenomenologically by a spin-flipping tunneling term in the Hamiltonian. The presence of a spin-flipping tunneling (SF) term, apart the usual spin-preserving (SP) one, has deep consequences on the interferometric properties of the set-up. In fact, if the spin is preserved during tunneling, the electron reverses its direction of propagation giving rise to "Fabry-P\'{e}rot" (F-P) interference phenomenon when traveling along the c
 losed loop shown in Fig. \ref{paths}, while if the electron flips its spin during the tunneling, it keeps moving in the same direction giving rise to "Aharonov-Bohm" (A-B) interference phenomenon, where the role of the F-P or A-B flux is played by the sum or the difference of the external gate voltages coupled to top and bottom edges. \cite{Dolcini11} Let us note that the A-B-like effect is absent in QH interferometers, since electrons are polarized by strong magnetic fields. Moreover, differently from QH interferometers, in QSH ones, F-P and A-B effects
can be controlled independently, permitting to generate pure charge or pure spin currents by all electrical means.\cite{Dolcini11} \\
Two important issues which have not been widely investigated in such interferometer setups are the role and strength of interactions within edge channels
and the role of the local spin-orbit interaction which determine the SF contributions to the current. The novelty of this work relies indeed on this twofold extension of previous analysis. From one side, in comparison with Ref. \onlinecite{Virtanen11} we consider a fully electrical set-up which is very versatile and allows to analyze the F-P and the A-B contribution to the current in a totally independent way avoiding possible drawback related to the presence of magnetic field. Moreover, in comparison with Refs.\onlinecite{Romeo12, Citro11} we investigate the interacting case that reveals by far not trivial at all and requires the introduction of totally different techniques with respect to the scattering matrix approach previously taken into account.\\
To address these questions, in our work we propose to realize an interacting quantum pump\cite{Thouless83} with the interferometric setup of Fig.\ref{fig1} where the time-dependent modulation of the point contacts potentials allows to change the ratio between the SP and the SF contribution to the current. Differently from other proposals (see Ref. \onlinecite{Krueckl11}), the relative strength of the two contributions to tunneling currents is not determined by the constriction geometry, but can be conveniently controlled by tuning the forcing signals. One of the effects we find is the different behavior, with respect to the pumping frequency, of two modulating functions characterizing the spin-preserving and the spin-flipping contribution to the current.  Moreover by means of the pump, spin currents are generated and controlled fully electrically. We will show that the behavior of the spin current, in particular the period of oscillation and the position of the zeros, is determined by the strength of the interactions, as well as the values of the F-P and A-B phases. This will allow us to find a way to extract information about the strength of interactions from the analysis of interference patterns.
 Since the setup is full electrically controllable, the study of the effects of the interactions can be carried out in a variety of different scenarios that could be investigated in future experiments.\\
The paper is organized as follows. In Sec. \ref{model} we introduce the helical Luttinger liquid description for the edge states of the 2D TIs. We describe the two QPCs geometry in which a DC spin current can be generated by means of periodically driven tunneling amplitudes out of phase. In Sec. \ref{Analytical_res} we analyze the SP and SF contributions to the spin current. The universality of the pumped spin charge as a function of the interactions is also discussed. Section \ref{Results} contains the analysis of the patterns of the spin current generated by F-P and A-B interference-like effect.
These are shown to be strongly dependent on the strength of the electron-electron interactions along the edges, thus allowing to extract information about the interactions themselves. Section \ref{conclusion} is devoted to conclusions.

\section{Model} \label{model}

\begin{figure}[h]
\centering
\includegraphics[scale=0.30]{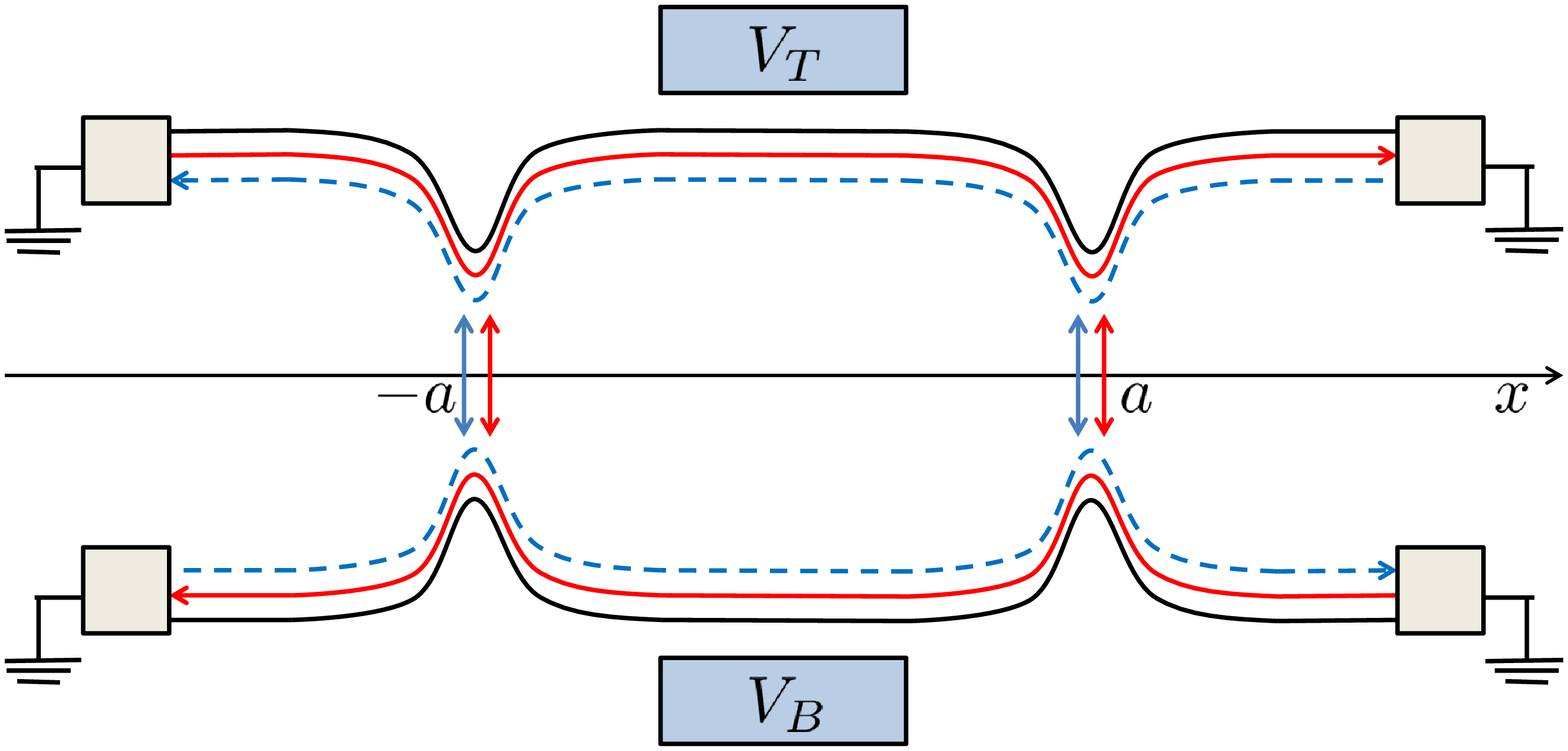}
\caption{(Color online) Top view of the four terminal interferometric setup. On the top edge one has right-moving spin up electrons (full line) and left-moving spin down electron (dashed line). The opposite holds for the bottom edge. The amplitudes of both SP (red) and SF (blue) electron tunneling at the QPCs placed at $x=\pm a$ are modulated in time with a frequency $\omega$ and phase difference $\theta$. The presence of the top ($V_{T}$) and bottom ($V_{B}$) gates can lead to a mismatch between the Fermi levels of the two edges ($k^{(T)}_{F}\neq k^{(B)}_{F}$). Shaded grey areas indicate the unbiased terminals.}
\label{fig1}
\end{figure}

\begin{figure}[h]
\centering
\includegraphics[scale=0.4]{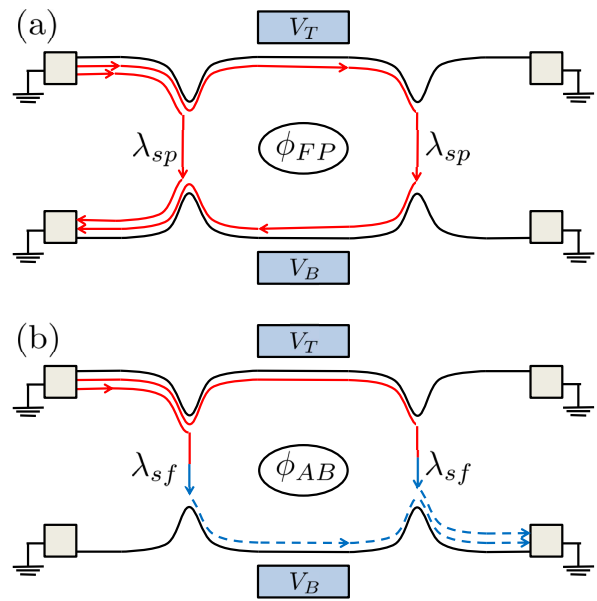}
\caption{(Color online) Two possible closed paths which correspond to (a) Fabry-P\'{e}rot and (b) Aharonov-Bohm interference effect induced by SP and SF tunneling events, respectively.}
\label{paths}
\end{figure}

We consider the QSH system in  a four-terminal geometry shown in Fig. \ref{fig1}, with a single Kramers doublet of helical edge states on each edge. Due to their helical properties \cite{Wu06} one can consider right-moving spin up and left-moving spin down electrons on the top edge ($T$) and the opposite on the bottom edge ($B$) (see Fig \ref{fig1}).
The corresponding Hamiltonians for free electrons are \cite{Hou09, Strom09}
\beq
\mathcal{H}_{T(B)}&=&-i\hbar v_{F} \int dx  \left(:\Psi^{\dagger}_{R, \uparrow (\downarrow)} \partial_{x}\Psi_{R, \uparrow (\downarrow)}\right. \nonumber \\
&&\left. - \Psi^{\dagger}_{L, \downarrow (\uparrow)} \partial_{x} \Psi_{L, \downarrow (\uparrow)}:\right)
\eeq
where $\Psi_{R, \uparrow}$ ($\Psi_{L, \uparrow}$) annihilates right (left)-moving electron with spin up, and analogous for the spin down, and $v_{F}$ is the Fermi velocity. With $:...:$ we indicate the normal ordering.

Concerning electron-electron interactions, the only terms that preserve the time-reversal symmetry of the system\cite{Moore06} are the dispersive
\beq
\label{perp}
\mathcal{H}_{d}&=& g_{2 \perp}\int dx:\left( :\Psi^{\dagger}_{R, \uparrow} \Psi_{R, \uparrow}::\Psi^{\dagger}_{L, \downarrow} \Psi_{L, \downarrow}:: \right. \nonumber \\
&& \left.+:\Psi^{\dagger}_{L, \uparrow} \Psi_{L, \uparrow}::\Psi^{\dagger}_{R, \downarrow} \Psi_{R, \downarrow}:\right):
\eeq
and the forward scattering
\be
\label{parallel}
\mathcal{H}_{f}=\frac{g_{4 \parallel}}{2} \sum_{\nu=R, L; \sigma=\uparrow, \downarrow} \int dx :\left(:\Psi^{\dagger}_{\nu, \sigma} \Psi_{\nu, \sigma}::\Psi^{\dagger}_{\nu, \sigma} \Psi_{\nu, \sigma}:\right):.
\ee
\newline
We will neglect possible Umklapp terms, which are relevant only at certain commensurate fillings and at very intense repulsive interaction strength. \cite{Wu06}

Through the standard bosonization technique \cite{Giamarchi03, Miranda03} one can write the electronic operators as
\beq
\Psi_{R,\uparrow}(x)&=&\frac{\mathcal{F}_{R,\uparrow}}{\sqrt{2 \pi \alpha}}e^{i k^{(T)}_{F}x} e^{-i\sqrt{2\pi} \varphi_{R,\uparrow}(x)}
\label{bosonized1}\\
\Psi_{L,\downarrow}(x)&=&\frac{\mathcal{F}_{L,\downarrow}}{\sqrt{2 \pi \alpha}}e^{-i k^{(T)}_{F}x} e^{-i\sqrt{2\pi} \varphi_{L,\downarrow}(x)}\\
\Psi_{R,\downarrow}(x)&=&\frac{\mathcal{F}_{R,\downarrow}}{\sqrt{2 \pi \alpha}}e^{i k^{(B)}_{F}x} e^{-i\sqrt{2\pi} \varphi_{R,\downarrow}(x)}\\
\Psi_{L,\uparrow}(x)&=&\frac{\mathcal{F}_{L,\uparrow}}{\sqrt{2 \pi \alpha}}e^{-i k^{(B)}_{F}x} e^{-i\sqrt{2\pi} \varphi_{L,\uparrow}(x)}
\label{bosonized4}
\eeq
with $\varphi_{R/L,\sigma}(x)$ ($\sigma=\uparrow, \downarrow$) bosonic fields, $\mathcal{F}_{R/L,\sigma}$ Klein factors necessary to give the proper commutation relations between electrons belonging to different edges, $\alpha$ a finite length cut-off;  $k^{(T)}_{F}$ and $k^{(B)}_{F}$ are the
Fermi momenta on the top and bottom edge respectively. Their expression in the presence of the applied gate voltages $V_T$ and $V_B$ is
$k_F^{(T,B)}=k_F+\kappa_{T,B}$  being $\kappa_{T,B}=K \frac{eV_{T,B}}{\hbar v_{F}}$ (see below for a precise definition of the Luttinger parameter $K$). Note that $k_F^{(T,B)}$ reduce to the non-interacting ones\cite{Dolcini11, Citro11} for $K=1$, as expected.

The bosonic fields $\varphi_{R/L,\sigma}(x)$ are related to the electron density along the edges through \cite{Giamarchi03, Miranda03}
\be
\rho_{R/L,\sigma}(x)=\mp\frac{1}{\sqrt{2\pi}}\partial_x\varphi_{R/L,\sigma}(x).
\ee

It is useful to introduce the helical basis for the bosonic fields on the upper and lower edge \cite{Giamarchi03, Miranda03, Virtanen11}
\be
\varphi_{T(B)}(x)=\frac{1}{\sqrt{2}}\left [\varphi_{L, \downarrow(\uparrow)}(x)-\varphi_{R, \uparrow(\downarrow)}(x)\right ],
\ee
with the canonically conjugated fields
\be
\Theta_{T(B)}(x)=\frac{1}{\sqrt{2}}\left [\varphi_{L, \downarrow(\uparrow)}(x)+\varphi_{R, \uparrow(\downarrow)}(x)\right ].
\ee
The low energy Hamiltonian assumes the typical form of an helical Luttinger liquid \cite{Hou09, Strom09, Dolcetto12, NoteH}
\be
\mathcal{H}_{eff}=\frac{v}{2}\sum_{i=T, B} \int dx \left[\frac{1}{K}\left(\partial_{x} \varphi_{i}\right)^{2}+K \left(\partial_{x}\Theta_{i}\right)^{2}\right],
\label{Heff}
\ee
with $K=\sqrt{\frac{2 \pi  v_{F}  \hbar +g_{4\parallel}-g_{2\perp}}{2\pi v_{F} \hbar+g_{4\parallel}+g_{2\perp}}}$ the interaction strength and $v= v_{F}\sqrt{\left(1+\frac{g_{4\parallel}}{2 \pi v_{F} \hbar}\right)^{2}-\left(\frac{g_{2\perp}}{ 2\pi v_{F} \hbar}\right)^{2}}$ the renormalized velocity.
In the following we will consider a pure Coulomb repulsion $(g_{4\parallel}=g_{2 \perp}>0)$ with $v=v_{F}/K$ and $K<1$. In order to avoid possible gapping effects associated to the Umklapp terms \cite{Wu06} or the presence of additional multiple electrons tunneling contribution \cite{Teo09, Schmidt11} we restrict our analysis in the interval $1/\sqrt{3} \leq K <1$.

Tunneling events become allowed when the top and bottom edge states are pinched through two QPCs as shown in Fig. \ref{fig1}. The only electron tunneling Hamiltonians that preserve time-reversal symmetry are \cite{Citro11, Note1}:

\be
\mathcal{H}^{sp}= \sum_{\sigma=\uparrow, \downarrow} \hbar  \int_{-\infty}^{+\infty} \!\!\!\! dx  \left\{\gamma_{sp}(x) \Psi^{\dagger}_{R, \sigma}(x) \Psi_{L, \sigma}(x)\right\}+\mathrm{h.c.},
\label{H_sp}
\ee
and ($\xi_{R, L}=\pm$)
\be
\mathcal{H}^{sf}= \sum_{\nu=R, L} \hbar \xi_{\nu}   \int_{-\infty}^{+\infty} \!\!\!\! dx \left\{\gamma_{sf}(x) \Psi^{\dagger}_{\nu, \uparrow}(x) \Psi_{\nu, \downarrow}(x) \right\}+ \mathrm{h.c.},
\label{H_sf}
\ee
which represent the SP and SF contribution to tunneling respectively. The latter term is strictly zero for inversion symmetric systems due to spin conservation, however the presence of local gates and deformations of the edges at the QPCs  locally can induce a spin-orbit coupling leading to SF processes. This fact has been already investigated both starting from the model proposed by Bernevig \emph{et al.} in Ref. \onlinecite{Bernevig06b} for the band structure \cite{Krueckl11} and in terms of  phenomenological descriptions. \cite{Teo09, Citro11, Dolcini11, Virtanen11, Vayrynen11}

In order to generate a DC spin current in absence of external bias one needs time-dependent tunneling amplitudes. They can be rendered time-dependent by periodically driven local electrostatic gates placed at the location of the QPCs,  $x=\pm a$:

\be
\gamma_{j}(x, t)= \lambda_{j} \left[\delta(x+a) \cos (\omega t +\theta) + \delta(x-a) \cos (\omega t)\right]
\label{tunneling_amplitude}
\ee
with $\lambda_{j}$ $(j=sp, sf)$ real parameters according to the requirement of time-reversal invariance, $\omega$ the pumping frequency and $\theta$ a phase shift assumed for simplicity equal for both SP and SF term.\cite{note-prob}\\
In the following we will calculate the spin current by a perturbative expansion up to the first order in the tunneling Hamiltonians $\mathcal{H}^{sp}$ and $\mathcal{H}^{sf}$ (second order in $\lambda_{sp}$ and $\lambda_{sf}$).

The last term in the total Hamiltonian is given by the coupling to the top and bottom gates, which reads:
\beq
H_g &=& \int_{-a}^{a} dx \,\,\,\, eV_T \left[ -\partial_x \varphi_{R \uparrow}(x)+\partial_x \varphi_{L \downarrow}(x)  \right] \nonumber \\
&+& eV_B \left[ -\partial_x \varphi_{R \downarrow}(x)+\partial_x \varphi_{L \uparrow}(x)  \right].
\label{coupling_gates}
\eeq
The gate voltages $V_T$ and $V_B$ shift the electronic spectrum and their difference breaks the degeneracy between top and bottom boundaries leading to different values of the Fermi momenta $k_F^{(T,B)}$.

\section{Spin current}\label{Analytical_res}

We focus the attention on the pumped spin current flowing in the system, defined as \cite{Citro11}
\be
I^{spin}(t)=q_s \left(\dot{N}_{R, \uparrow}-\dot{N}_{L,\downarrow}\right)
\label{spin_current}
\ee
where $q_s$ is the quantum of spin $\hbar/2$ and
\be
N_{\nu, \sigma}= \int_{-\infty}^{+\infty} d x :\Psi^{\dagger}_{\nu, \sigma} (x) \Psi_{\nu, \sigma} (x):.
\ee
The charged current flowing into the system can be obtained in an analogous way by exploiting the general relations derived in the so called XYZ decomposition \cite{Teo09} that holds for helical edge states in a four-terminal setup.
The spin current is given by two contributions coming from the tunnelings (\ref{H_sp}) and (\ref{H_sf}):
\be
I^{spin}(t)=I^{sp}(t)+I^{sf}(t)
\label{spin_current}.
\ee
The former
\be
I^{sp}(t)=-i q_s \sum_{\sigma=\uparrow, \downarrow} \int_{-\infty}^{+\infty} dx \gamma_{sp}(x, t) \Psi^{\dagger}_{R, \sigma}(x) \Psi_{L, \sigma}(x) + \mathrm{h. c.}
\ee
is associated to SP tunneling, while the latter
\be
I^{sf}(t)=-i q_s \sum_{\nu=R, L} \xi_{\nu} \int_{-\infty}^{+\infty} dx \gamma_{sf}(x, t) \Psi^{\dagger}_{\nu, \uparrow}(x) \Psi_{\nu, \downarrow}(x)+\mathrm{h. c.}
\ee
is associated to SF tunneling.

The expectation value of the spin current in Eq. (\ref{spin_current}) is given, according to the Keldysh contour formalism (see Refs. \onlinecite{Citro03, Martin05, Kamenev09} and references therein for detailed discussions), by

\be
\langle I^{spin}(t) \rangle =\sum_{j=sp, sf} \langle I^{j}(t) \rangle
\ee
with

\be
\langle I^{j}(t) \rangle=\frac{1}{2} \sum_{\eta = \pm } \langle \mathcal{T}_{K} \left\{ I^{j}(t^{\eta}) e^{-i \int _{\mathcal{C}} dt_{1} \mathcal{H}^{j} (t_{1})} \right\}  \rangle
\ee
where all the operators have to be considered in the interaction picture with respect to $\mathcal{H}_{eff}$ in Eq. (\ref{Heff}) and we put $\hbar=1$. The time-ordering along the Keldysh contour $\mathcal{C}$ is indicated by $\mathcal{T}_{K}$ and $\eta=\pm$ labels the upper and the lower branch of the contour respectively.

At the first order in the tunneling Hamiltonians one has the contribution
\begin{equation}\label{SpinCurrent}
\langle I^{j}(t) \rangle=-\frac{i}{2} \sum_{\eta, \eta_{1}=\pm} \eta_{1} \int_{-\infty}^{+\infty} d t_{1} \langle T_{K} I^{j}(t^{\eta}) \mathcal{H}^{j}(t^{\eta_{1}}_{1})\rangle
\end{equation}
where the thermal average is taken with respect to the ground state of $\mathcal{H}_{eff}$.
Let us note that the main contributions to  $\langle I^{sp}(t) \rangle$ and $\langle I^{sf}(t) \rangle$  come from the interference effects of the wavefunctions (\ref{bosonized1})-(\ref{bosonized4}) traveling along closed paths between the two QPCs. When only SP tunneling is taken into account the interference will be of F-P type and the phase acquired by the electron along the path is $\phi_{\mathrm{FP}}=2\left(k^{(T)}_{F}+ k^{(B)}_{F}\right)a\propto \left (V_T+ V_B\right )$, while if only SF tunneling is considered the interference will be of A-B type and the phase acquired by the electron will be
$\phi_{\mathrm{AB}}=2\left(k^{(T)}_{F}- k^{(B)}_{F}\right)a\propto \left (V_T- V_B\right )$. Some of the closed paths which give raise to F-P and A-B interference and their link to the tunneling processes, SP or SF, are shown in Fig. \ref{paths}.\\
Let us also note that, given the tunneling time-dependence in Eq. (\ref{tunneling_amplitude}), the spin-current will include a DC contribution (pumped current) and an AC contribution of frequency $2\omega$.
In the following we will focus on the DC component only. In this case one has
\be
\langle I^{spin}_{DC} (\omega) \rangle= \langle I^{sp}_{DC} (\omega) \rangle+\langle I^{sf}_{DC} (\omega) \rangle
\label{totspin}
\ee
where
\beq
\langle I^{sp}_{DC} (\omega) \rangle&=&\frac{i q_s \lambda_{sp}^{2}}{\pi^{2} \alpha^{2}} \sin{\phi_{\mathrm{FP}}} \sin{\theta} \nonumber\\
&&\int_{-\infty}^{+\infty}\!\! d \tau \sin(\omega \tau) e^{\left[ \mathcal{W}_{R}(2a, \tau) +\mathcal{W}_{L}(2a, \tau)\right]}
\label{Isp_DC}
\eeq
and
\beq
\langle I^{sf}_{DC} (\omega) \rangle&=&\frac{i q_s \lambda_{sf}^{2}}{2\pi^{2} \alpha^{2}} \sin{\phi_{\mathrm{AB}}} \sin{\theta} \nonumber \\
&&\sum_{\nu=R, L}\int_{-\infty}^{+\infty} \!\!d \tau \sin(\omega \tau) e^{2 \mathcal{W}_{\nu}(2a, \tau)}
\label{Isf_DC}
\eeq
where $\phi_{\mathrm{FP/AB}}$ are the F-P and A-B like phases\cite{Dolcini11, Romeo12, Citro11}
that can be controlled independently, simply by varying the external gate voltages.
It is worth to note that a non zero phase difference in the time dependence of the tunneling amplitudes of the two QPCs  ($\theta \ne 0$) is crucial in order to have a DC component of the spin current. This is in agreement with the general statements of the parametric pumping. \cite{Thouless83}

The bosonic Green's functions that appear in the Kernel of the integrals above are \cite{Ferraro10, Carrega11, Carrega12}

\begin{eqnarray}
\mathcal{W}_{R/L}(x,t)&=&2\pi\langle \left[\varphi_{R/L, \sigma}(x, t)- \varphi_{R/L, \sigma}(0, 0)\right]\varphi_{R/L, \sigma}(0,0)\rangle \nonumber \\
&=&c^{(+/-)}_K \mathcal{W}_{+}(x,t)+c^{(-/+)}_K\mathcal{W}_{-}(x,t)
\end{eqnarray}
with
\begin{equation}
\mathcal{W}_{\pm}(x,t)=\mathcal{W}\left(t\mp \frac{x}{v} \right)
\end{equation}
and
\begin{equation}
\mathcal{W}(t)=\ln \left [\frac{\left |\Gamma\left (1+\frac{k_{B}T}{\omega_{c}}-ik_{B}T t \right )\right |^2}{\Gamma^2\left (1+\frac{k_{B} T}{\omega_{c}}\right )\left (1+i\omega_{c} t\right )}\right ].
\label{W_exact}
\end{equation}
Here
\be
c^{(\pm)}_K=\frac{1}{4}\left (\sqrt{K}\pm\frac{1}{\sqrt{K}}\right )^2
\ee
are the interaction dependent coefficients, $\Gamma(x)$ is the Euler Gamma function \cite{Gradshteyn94}, $T$ the temperature  and $\omega_{c}=v/\alpha$ the energy bandwidth. This latter quantity fixes the limit of validity of the helical Luttinger liquid description and can be related to the energy gap between the bulk conduction and valence bands of the heterostructure which, in realistic experimental setup \cite{Konig07, Hou09, Konig08}, is $\Delta \approx 30$ meV. According to this assumption one has $\omega_{c}\sim \Delta/\hbar \approx 50$ THz that represents by far the greatest frequency involved in the problem.\\
We need now to explicitly evaluate the integrals in Eqs. \eqref{Isp_DC} and \eqref{Isf_DC}. Analogous integrals have already been calculated in Refs. \onlinecite{Chamon97, Virtanen11}, therefore in the following we only report the main results, while the key points of the derivation are recalled in Appendix A (zero temperature case) and B (finite temperature case).\\
At zero temperature the SP and SF current contributions can be written respectively as
\be
\langle I^{sp}_{DC}\rangle =  \frac{\lambda^{2}_{sp}q_s}{2 \pi^{2} \alpha^{2}} \sin{\theta}\sin{\phi_{\mathrm{FP}}} H^{(0)}\left(d_{K}, K \omega/\omega_{0}\right)\tilde{\mathcal{P}}^{(0)}_{2d_{K}}(\omega)
\label{I_zeroDCt}
\ee
and
\beq
\langle I^{sf}_{DC} \rangle&=& \frac{\lambda^{2}_{sf}q_s}{2 \pi^{2} \alpha^{2}} \sin{\theta}\sin{\phi_{\mathrm{AB}}} \nonumber \\
&& J^{(0)}\left(2c^{(+)}_{K}, 2c^{(-)}_{K}, K \omega/\omega_{0}\right)\tilde{\mathcal{P}}^{(0)}_{2d_{K}}(\omega),
\label{I_sft}
\eeq
where $H^{(0)}$ and $J^{(0)}$ are proper modulating functions whose explicit form is derived in Appendix A and $\mathcal{P}^{(0)}_{\alpha}$ is the zero temperature electronic Green's function in the energy space (see Eq. (\ref{P_zero})).\\
Analogously for the finite temperature case one has (see Appendix B)
\beq
\langle I^{sp}_{DC}\rangle &=&  \frac{\lambda^{2}_{sp}q_s}{2 \pi^{2} \alpha^{2}} \sin{\theta}\sin{\phi_{\mathrm{FP}}} H\left(d_{K}, K \omega/\omega_{0},K k_{B}T/\omega_{0}\right)\nonumber \\
&& \left[\tilde{\mathcal{P}}_{2d_{K}}(\omega, T)-\tilde{\mathcal{P}}_{2d_{K}}(-\omega, T)\right]
\eeq
and
\beq
\langle I^{sf}_{DC} \rangle&=& \frac{\lambda^{2}_{sf}q_s}{2 \pi^{2} \alpha^{2}} \sin{\theta}\sin{\phi_{\mathrm{AB}}} \nonumber \\
&& J\left(2c^{(+)}_{K}, 2c^{(-)}_{K}, K \omega/\omega_{0}, K k_{B}T/\omega_{0}\right) \nonumber \\
&&\left[\tilde{\mathcal{P}}_{2d_{K}}(\omega, T)-\tilde{\mathcal{P}}_{2d_{K}}(-\omega, T)\right].
\label{I_sf}
\eeq
where now both the modulating functions $H$, $J$ and the electronic Green's function $\mathcal{P}_{\alpha}$ are evaluated at non zero temperature $T$.\\
Note that on the above equations we introduced the short notations
\be
d_{K}=c^{(+)}_{K}+c^{(-)}_{K}= \frac{1}{2}\left( K+ \frac{1}{K}\right)
\label{d}
\ee
and $\omega_{0}=v_{F}/2a$.

\subsection{Universality of the pumped spin}

The pumped spin charge is obtained by integrating the DC spin current in Eq. (\ref{totspin}) over a period, i.e. $Q_s=2\pi \langle I^{spin}_{DC}\rangle/\omega$. One can immediately see that, perturbatively, the spin charge pumped in a period is proportional to the area in the parameter space encircled by the pumping parameters, i.e. $q_s\lambda_{i}^2 \sin \theta$ ($i=sp,sf$) and is a non-universal quantity. A universal behavior for the pumped spin charge in a cycle can be inferred when linking it to the renormalization fixed points of the helical LL with a single impurity.\cite{Sharma01, Sharma03, note_universality} The main argument is the following.
The spin  current originates from the non-equilibrium SP and SF contribution, due to the explicit time
dependence in the tunneling Hamiltonian in Eq. (\ref{H_sp}-\ref{H_sf}). This is
in contrast to the DC current generated by a source-drain voltage: $I = I_{d} - I_{b}$. In this case there are two distinct contributions to the current: the direct one $I_{d}$, arising from the applied DC voltage and a backscattered part $I_{b}$. In the case of a quantum pump, no source-drain voltage is applied, so there is no direct contribution ($I_{d }= 0$). Therefore, the DC pumping current arises from the backscattering term (in our case both spin-preserving and spin-flipping backscattering are contributing): $I_{p} = -I_{b}$. To have a quantized spin charge pumped in a cycle, one requires that backscattering becomes a relevant operator, driving the system to an insulating fixed point. Since a relevant backscattering is equivalent to an irrelevant tunneling, given the renormalization group flow for the spin-flip and spin-preserving tunneling $\frac{d \lambda_{i}}{dl}=\left( 1-(K^{-1}+K)/2\right)\lambda_{i}$, the condition is satisfied whenever  $(K^{-1}+K)/2>1$. The res
 ult is a perfectly quantized spin charge and the quanta are given by the winding number of the total backscattering amplitude.

\section{Results} \label{Results}
\subsection{Modulating functions}

Before entering into the details of the behavior of the spin current, it is useful to analyze the peculiar forms of the modulating functions both at zero and finite temperature whose functional expressions are reported in Appendices A and B. Fig. \ref{fig2} (a) shows $H^{(0)}$ for different values of the interaction strength $K$ as a function of $\omega/\omega_{0}$. As already discussed in Ref. \onlinecite{Chamon97}, in the framework of the interferometric properties of the fractional quantum Hall effect, it shows an oscillating and rapidly decaying behavior as a function of the pumping frequency due to the presence of the Bessel function. These oscillations are related to the finite distance between the QPCs. For very close point contacts this behavior disappears.  The envelope of the curves decreases algebraically as $(\omega/\omega_{0})^{-\frac{1}{2}(K+1/K)}$ (see Eq. (\ref{H})). The zeros of this function are given  approximatively by
\be
\omega/\omega_{0}\approx \frac{\pi}{K}\left[n+ \frac{(1+d_{K})}{2}\right] \qquad n\in \mathbb N
\ee
moving to higher frequencies by decreasing $K$. Consequently the number of observed oscillations in the considered range of frequencies decrease by increasing the interaction.

An analogous behavior can be found for the modulating function $J^{(0)}$ (see Fig. \ref{fig2} (b)) that is related to the SF tunneling processes. Here, the zeros can be found at
\be
\label{Jzeros}
\omega/\omega_{0}\approx \frac{\pi}{K}\left(\frac{d_{K}}{2}+n\right)
\ee
and, in the considered range of interactions $0.6\leq K\leq 1$, the oscillations are only slightly suppressed at high frequency, with an envelop given by $(\omega/\omega_{0})^{1-\frac{1}{2}(K+1/K)}$.

\begin{figure}[h]
\centering
\includegraphics[scale=0.38]{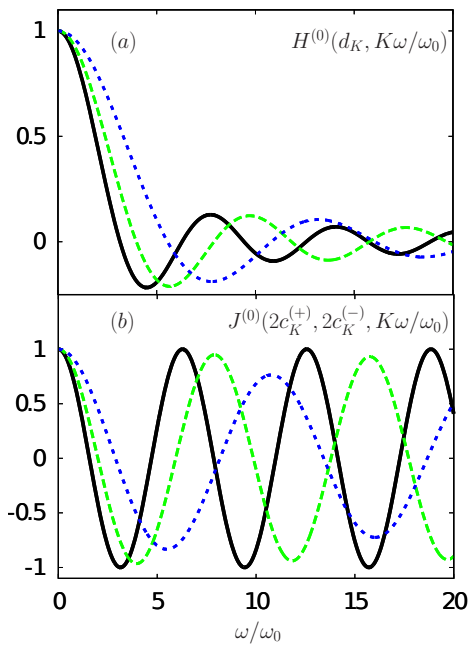}
\caption{(color online) Modulating functions (a) $H^{(0)}(d_{K}, K \omega/\omega_{0})$ and (b) $J^{(0)}(2 c^{(+)}_{K}, 2 c^{(-)}_{K}, K \omega/\omega_{0})$ as a function of $\omega/\omega_{0}$ for $K=1$ (non-interacting case, full black curve), $K=0.8$ (dashed green curve) and $K=0.6$ (short dashed blue curve).}
\label{fig2}
\end{figure}

\begin{figure}[h]
\centering
\includegraphics[scale=0.38]{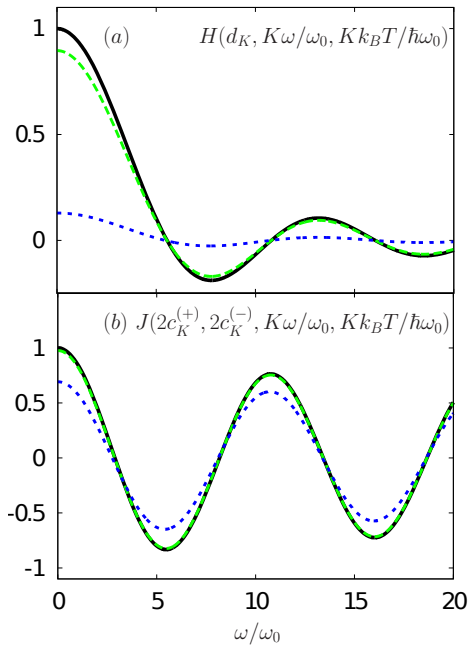}
\caption{(Color online) Modulating functions as a function of $\omega/\omega_{0}$ for $K=0.6$ at various temperatures: $k_{B}T=0$ (full black curve), $k_{B}T=0.2 \omega_{0}$ (dashed green curve) and $k_{B}T=\omega_{0}$ (short dashed blue curve). (a) $H(d_{K}, K \omega/\omega_{0}, K k_{B}T/\omega_{0})$ and (b) $J(2 c^{(+)}_{K}, 2 c^{(-)}_{K}, K \omega/\omega_{0}, K k_{B}T/\omega_{0})$.}
\label{fig3}
\end{figure}

The comparison with the finite temperature case is shown in Fig. \ref{fig3} for $K=0.6$ and $T\leq T_0=\omega_0/k_B$ (analogous results can be found for other values of the interaction). It is possible to note that oscillations of $H$ in Eq. (\ref{H_temperature}) (see Fig. \ref{fig3} (a))  are progressively suppressed by increasing the temperature. However, the periodicity and the zeroes positions are not affected. The function $J$ in Eq. (\ref{J_temperature}) is only slightly affected by the temperature with respect to $H$, as can be seen in Fig. \ref{fig3} (b), so spin-flipping tunneling terms are quite robust and dominate the transport at high enough frequency and over a large temperature range. This is a consequence of the differences in the interference paths induced by the helical properties of the edge states (see Fig. \ref{paths}). \cite{Dolcini11, Citro11, Virtanen11} For $T\gg T_{0}$ (temperature regime that is out of the aim of this paper) the features of both modu
 lating functions are washed out and additional tunneling processes need to be taken into account (see Ref. \onlinecite{Virtanen11} for a detailed discussion of this regime). Note that at very low frequency ($\omega \ll \omega_{0}$), the effects of the modulating functions becomes less important $(H, J \approx 1)$, although one notices a dependence of the concavity on the interaction $K$. \cite{Sharma01} \\
Assuming $a\approx 1$ $\mu$m and $v_{F}\approx  10^5$ m/s, as appears  from experiments \cite{Konig07, Hou09, Strom09, Konig12}, one obtains $\omega_{0} \approx 50$ GHz\cite{Note3b} and consequently $T_{0}$ of the order of hundred mK that have to be compared with the typical temperature at which experiments are carried out. \cite{Konig07, Roth09}

\subsection{Current}
\begin{figure}[h]
\centering
\includegraphics[scale=0.40]{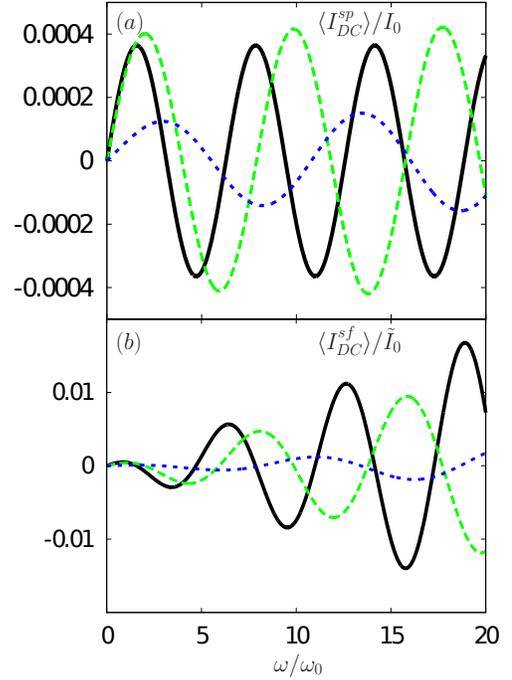}
\caption{(Color online) (a) SP and (b) SF contributions to the DC spin current  (in units of $I_{0}=\frac{q_s \lambda^{2}_{sp}}{2 \pi^{2} \alpha^{2} \omega_{c}} \sin{\theta}$ and $\tilde{I}_{0}=\frac{q_s \lambda^{2}_{sf}}{2 \pi^{2} \alpha^{2} \omega_{c}} \sin{\theta}$ respectively) as a function of $\omega$ (in units of $\omega_{0}$) for: $K=1$ (non-interacting case, full black curves), $K=0.8$ (dashed green curves) and $K=0.6$ (short dashed blue curves). Other parameters are: $\omega_{0}/\omega_{c}=2 \times 10^{-4}$, $\phi_{\mathrm{FP}}=\phi_{\mathrm{AB}}=\pi/4$ $(\mathrm{mod}\,\,\,\, 2\pi)$ and $T/T_{0}=0.4$.}
\label{fig4}
\end{figure}

The evolution of the SP and SF current contributions at finite temperature, as a function of the pumping frequency $\omega$, is shown in Fig. \ref{fig4} for different values of interactions. The behavior is dominated by the power-law behavior typical of helical Luttinger liquids $\tilde{\mathcal{P}}^{(0)}_{2d_{K}}(\omega)\propto (\omega/\omega_{c})^{K+\frac{1}{K}-1}$, already predicted for the current as a function of the source-drain bias \cite{Strom09, Dolcetto12}, strongly modulated by the oscillating functions $H$ (Fig. \ref{fig4} (a)) and $J$ (Fig. \ref{fig4} (b)) respectively.
The different envelop power law between $H^{(0)}$ and $J^{(0)}$ (cf. Fig. \ref{fig2}) creates a qualitatively different behavior between the SP and SF contributions to the current. Namely, even in the presence of moderate interactions ($K>0.6$) the SF contribution may dominate at high frequencies.

\begin{figure}[h]
\centering
\includegraphics[scale=0.57]{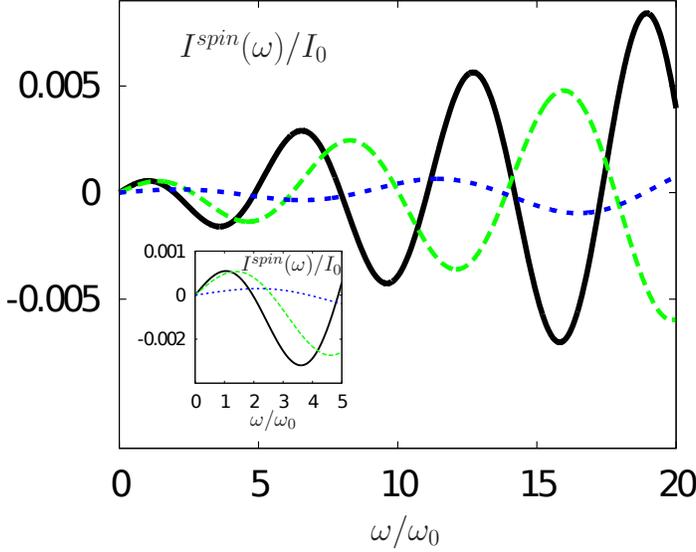}
\caption{(Color online) Spin current $I_{spin}(\omega)$ (in units of $I_{0}=\frac{q_s\lambda^{2}_{sp}}{2 \pi^{2} \alpha^{2} \omega_{c}} \sin{\theta}) $ as a function of $\omega$ (in units of $\omega_{0}$) for: $K=1$ (non-interacting case, full black curves), $K=0.8$ (dashed green curves) and $K=0.6$ (short dashed blue curves). Other parameters are: $\omega_{0}/\omega_{c}=2 \times 10^{-4}$, $\lambda_{sf}^{2}/\lambda_{sp}^{2}=0.5$, $\phi_{\mathrm{FP}}=\phi_{\mathrm{AB}}=\pi/4$ $(\mathrm{mod}\,\,\,\, 2\pi)$ and $T/T_{0}=0.4$ ($T_{0}=\hbar \omega_{0}/k_{B}$). The inset shows the physically relevant regime where the first zero of the current appears.}
\label{fig5}
\end{figure}

Fig. \ref{fig5} shows the total DC pumped spin current. According to the previous considerations, at high frequency (above some unity of $\omega_{0}$), this quantity is driven only by the SF contribution due to the algebraic damping of the SP one, therefore a detailed study of this regime can provide important information about the role played by this kind of processes in the dynamics of the system. Let us note that the physically relevant regime is the one up to the first zero of the current  and here two interesting features take place related to electron interactions: both the concavity and the location of the zero of the spin current depend on the interaction. A measurement of such characteristics could give information on the interaction strength itself. To be more precise the location of the zeroes of the current allows to extract the value of $K/\omega_{0}$ and this, together with independent measurements of the point contact distance $a$ and the Fermi velocity $v_{F}$, makes possible to obtain the value of $K$. In general the position of the zeroes is a complicated function that is affected both by SP and SF contribution but, by properly acting on the value of the F-P and A-B phase, it is possible to alternatively turn them on and off leading to a simplification.

\begin{figure}[h]
\centering
\includegraphics[scale=0.45]{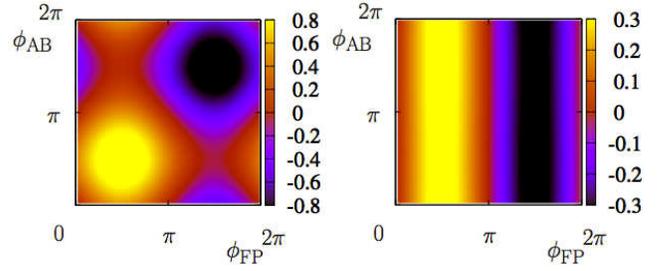}
\caption{(Color online) Density plot of the spin current $I_{spin}(\omega)$ (normalized with respect to $I_{point}(\omega, T)=\frac{q_s\lambda^{2}_{sp}}{2 \pi^{2} \alpha^{2}} \sin{\theta}\left[\tilde{\mathcal{P}}_{2d_{K}}(\omega, T)-\tilde{\mathcal{P}}_{2d_{K}}(-\omega, T)\right]$)  as a function of the Fabry-P\'{e}rot $\phi_{\mathrm{FP}}$ ($x$ axis) and Aharonov-Bohm $\phi_{\mathrm{AB}}$ ($y$ axis) phases for different pumping frequency: $\omega/\omega_{0}=0.4$ (left panel), $\omega/\omega_{0}=2$ (right panel).
Other parameters are: $T/T_{0}=0.4$, $\lambda_{sf}^{2}/\lambda_{sp}^{2}=0.5$ and $K=0.8$.}
\label{fig6}
\end{figure}

Even more interesting is the dependence of the current on the Fabry-P\'{e}rot and Aharonov-Bohm phase as a function of the pumping frequency and the interaction. As already shown in the non-interacting case one expects very intriguing patterns. \cite{Citro11} This is due to the different modulations in amplitude and sign of the SP and SF contributions to the spin current. Some examples of these patterns are shown in Fig. \ref{fig6} for various frequencies and at fixed interaction ($K=0.8$). At low pumping frequency (left panel) both the SP and SF contributions affect the current leading to clear periodicities both in $\phi_{\mathrm{FP}}$ and $\phi_{\mathrm{AB}}$.\cite{Note4} The amplitude of the oscillations depends on the ratio $\lambda_{sf}/\lambda_{sp}$. Usually in literature these tunneling amplitudes are assumed to have a comparable order of magnitude, with $\lambda_{sp}> \lambda_{sf}$. \cite{Teo09, Dolcini11, Citro11, Krueckl11, Liu11}

The bare parameters $\lambda_{sp,sf}$ characterize the QPC tunneling properties and are fixed by the geometry of the constriction when it is realized by using an etching procedure.
However the relative strength of $\lambda_{sf,sp}$ is dynamically renormalized by the ac driving (pumping) as can be deduced by the Eqs. (\ref{I_zeroDCt})-(\ref{I_sft}) where the renormalized quantities  $\lambda^{\ast}_{sp}=\lambda_{sp}\sqrt{H(...)}$, $\lambda^{\ast}_{sf}= \lambda_{sf}\sqrt{J(...)}$ can be recognized. The dynamical renormalization allows to change the ratio  $\lambda^{\ast}_{sf}/\lambda^{\ast}_{sp}$ compared to its zero-frequency value (i.e. $\lambda_{sf}/\lambda_{sp}$) allowing  to explore the full interference pattern. In this way the ratio $\langle I^{sp}(\omega) \rangle /\langle I^{sf} (\omega)\rangle$ can be controlled by acting on the pumping frequency. Thus the pumping-driven interferometer proposed here allows to amplify or deplete the effective value of $\lambda^{\ast}_{sf}$ assuming its bare value to be non-vanishing.

The comparison with measurements on interfering patterns will allow to extract the relative weights of the bare parameters $\lambda_{sp,sf}$.

\begin{figure}[!ht]
\centering
\includegraphics[scale=0.45]{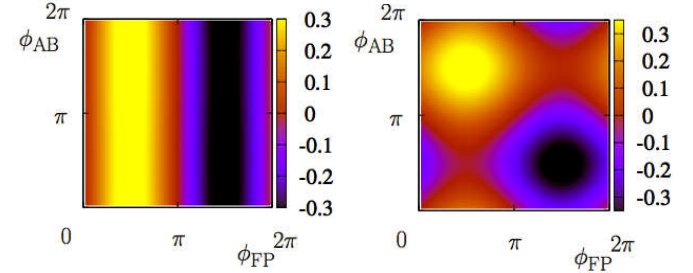}
\caption{(Color online) Density plot of the spin current $I_{spin}(\omega)$ (normalized with respect to $I_{point}(\omega)=\frac{q_s\lambda^{2}_{sp}}{2 \pi^{2} \alpha^{2}} \sin{\theta}\left[\tilde{\mathcal{P}}_{2d_{K}}(\omega, T)-\tilde{\mathcal{P}}_{2d_{K}}(-\omega, T)\right]$) as a function of the Fabry-P\'{e}rot $\phi_{\mathrm{FP}}$ ($x$ axis) and Aharonov-Bohm $\phi_{\mathrm{AB}}$ ($y$ axis) phases for interaction $K=0.8$ (left panel) and $K=1$ (right panel, non-interacting case). Other parameters are: $\omega/\omega_{0}=2$, $T/T_{0}=0.4$ and $\lambda_{sf}^{2}/\lambda_{sp}^{2}=0.5$.}
\label{fig7}
\end{figure}

According to the previous considerations, in this kind of setup it is therefore possible to fine tune the pumping frequency in such a way to make null the SF contribution to the spin current but still keeping the SP one finite; in this case it is possible to note a periodic pattern with respect to $\phi_{\mathrm{FP}}$ only (right panel of Fig.7). Due to the robustness of the modulating function $J$ associated to the SF contribution with respect to thermal effects and its quite negligible decay at high frequency, the experimental observation of vertically striped picture of Fig.\ref{fig6} represents a very precise measurement of the positions of the zeroes of the SF contribution (see Eq. \eqref{Jzeros}) and consequently of the value of the interaction strength among the electrons.
In the opposite case, when the SP contribution is null and the SF one is different from zero, one has horizontally striped patterns driven only by $\phi_{\mathrm{AB}}$ (not shown). \\
The dependence of the interference path on interaction strength along the edge can be seen in Fig.  \ref{fig7} where the interacting (left panel) and non-interacting (right panel) spin currents are shown at the same pumping frequency. While the left panel clearly shows a vertically striped pattern signature of a null SF contribution, the right one is more similar to the top left picture of Fig. \ref{fig6} because of the finite contribution of both the SF and the SP terms.

It is worth to note that one needs the presence of both SP and SF tunneling processes occurring at the QPCs in order to observe the rich variety of interference patterns described above, nevertheless the results remain qualitatively the same also when their relative strength is smaller then the one considered here.
From the above discussions, it follows that the presence of spin-flipping terms can be experimentally verified via interference patterns.

\section{Conclusions} \label{conclusion}

We proposed an interacting and parametric quantum pump for the helical edge states of a two-dimensional topological insulator in a four-terminal set-up. Here, a DC pumped spin current is induced by periodically modulating in time spin-preserving and spin-flipping electron tunneling amplitudes out of phase.  We analyzed the contributions of these processes to the spin current both at zero and finite temperature.  We found that the power-law behavior typical of the helical Luttinger liquid expected for the spin current in a single QPC is strongly affected by interference phenomena of Fabry-P\'{e}rot and Aharonov-Bohm type whose effect is encoded into modulating functions for the current. We investigated the interference patterns of the spin-current and its dependence on Fabry-P\'{e}rot and Aharonov-Bohm phases controlled by external electrical gates. The measurement of their shape, and in particular the locations of their zeros, represents an important tool in order to extract
information about the presence of spin-flipping terms and on the strength of the electron-electron interaction in the system.

Recent achievements in interferometric devices with integer quantum Hall states \cite{Fazio11} make us confident about the feasibility of a quantum pump for helical edges states in topological insulators.

\appendix
\section{Current contributions at zero temperature}\label{app:A}

At zero temperature, Eq. (\ref{W_exact}) reduces to
\be
\mathcal{W}^{(0)}(t)=-\ln \left (1+i\omega_{c} t\right ),
\ee
therefore the SP term to the current can be written as
\be
\langle I^{sp}_{DC}\rangle =  \frac{\lambda^{2}_{sp}q_s}{2 \pi^{2} \alpha^{2}} \sin{\theta}\sin{\phi_{\mathrm{FP}}} H^{(0)}\left(d_{K}, K \omega/\omega_{0}\right)\tilde{\mathcal{P}}^{(0)}_{2d_{K}}(\omega)
\label{I_zeroDC}
\ee
where

\be
d_{K}=c^{(+)}_{K}+c^{(-)}_{K}= \frac{1}{2}\left( K+ \frac{1}{K}\right)
\label{d}
\ee
and $\omega_{0}=v_{F}/2a$.

In Eq. (\ref{I_zeroDC}) appears the modulating function\cite{Chamon97, Chevallier10}
\be
H^{(0)}(\xi, x)= \sqrt{\pi} \frac{\Gamma(2 \xi)}{\Gamma(\xi)}\frac{\mathcal{J}_{\xi-\frac{1}{2}}(x)}{(2x)^{\xi-\frac{1}{2}}},
\label{H}
\ee
with $\mathcal{J}_{\xi-\frac{1}{2}}(x)$ Bessel function of the first kind \cite{Gradshteyn94} and
\be
\tilde{\mathcal{P}}^{(0)}_{\xi}(\omega)= \frac{2 \pi}{\Gamma(\xi)\omega_{c}}\left(\frac{\omega}{\omega_{c}}\right)^{\xi-1} e^{- \omega/\omega_{c}}\theta{(\omega)}
\label{P_zero}
\ee
is the Fourier transform of the zero temperature electronic Green's function $\mathcal{P}^{(0)}_{\xi}(t)= e^{\xi\mathcal{W}^{(0)}(t)}$.

In an analogous way the SF contribution reads
\beq
\langle I^{sf}_{DC} \rangle&=& \frac{\lambda^{2}_{sf}q_s}{2 \pi^{2} \alpha^{2}} \sin{\theta}\sin{\phi_{\mathrm{AB}}} \nonumber \\
&& J^{(0)}\left(2c^{(+)}_{K}, 2c^{(-)}_{K}, K \omega/\omega_{0}\right)\tilde{\mathcal{P}}^{(0)}_{2d_{K}}(\omega),
\label{I_sf}
\eeq
that depends on the other modulating function
\beq
J^{(0)}(\xi_{1}, \xi_{2}, x)&=&\frac{1}{2} \left[e^{-i x} {}_{1}F _{1} \left(\xi_{1}, \xi_{1}+\xi_{2}; 2i x\right) \right. \nonumber \\
&&\left. + e^{-i x} {}_{1}F _{1} \left(\xi_{2}, \xi_{1}+\xi_{2}; 2i x \right)\right]
\label{J}
\eeq
with ${}_{1}F_{1}(a, b; z)$ the Kummer Hypergeometric confluent function. \cite{Gradshteyn94} Note that the above function admits the limit $J^{(0)}\left(2, 0, x\right)= \cos{x}$  and  $J^{(0)}\left(\xi, \xi, x\right)=H^{(0)}\left(\xi, x\right)$.\cite{Note3}

\section{Current contributions at finite temperature}\label{app:B}

At finite temperature, as long as $k_{B}T \ll \omega_{c}$ and $\omega_{c} t \gg 1$, Eq. (\ref{W_exact}) reduces to

\be
\mathcal{W}(t)\approx  \frac{\pi k_{B}T t}{\sinh{\left(\pi k_{B}T t\right)}}.
\ee
Also in this case it is possible to write the SP and SF contributions to the spin current in a factorized way as
\beq
\langle I^{sp}_{DC}\rangle &=&  \frac{\lambda^{2}_{sp}q_s}{2 \pi^{2} \alpha^{2}} \sin{\theta}\sin{\phi_{\mathrm{FP}}} H\left(d_{K}, K \omega/\omega_{0},K k_{B}T/\omega_{0}\right)\nonumber \\
&& \left[\tilde{\mathcal{P}}_{2d_{K}}(\omega, T)-\tilde{\mathcal{P}}_{2d_{K}}(-\omega, T)\right]
\eeq

and
\beq
\langle I^{sf}_{DC} \rangle&=& \frac{\lambda^{2}_{sf}q_s}{2 \pi^{2} \alpha^{2}} \sin{\theta}\sin{\phi_{\mathrm{AB}}} \nonumber \\
&& J\left(2c^{(+)}_{K}, 2c^{(-)}_{K}, K \omega/\omega_{0}, K k_{B}T/\omega_{0}\right) \nonumber \\
&&\left[\tilde{\mathcal{P}}_{2d_{K}}(\omega, T)-\tilde{\mathcal{P}}_{2d_{K}}(-\omega, T)\right].
\label{I_sf}
\eeq

The finite temperature modulating functions are now\cite{Chamon97, Virtanen11}

\begin{widetext}

\be
H(\xi, x, y)= 2 \pi \frac{\Gamma{(2\xi)}}{\Gamma{(\xi)}}\frac{e^{-2\pi \xi y}}{\sinh \left(\frac{x}{2y}\right)}\Im \left\{ \frac{e^{i x}}{\Gamma{\left(\xi+i \frac{x}{2\pi y}\right)}\Gamma{\left(1-i \frac{x}{2 \pi y}\right)}} {}_{2} F_{1}\left(\xi, \xi-i\frac{x}{2\pi y}, 1-i \frac{x}{2\pi y}; e^{-4 \pi y}\right)\right\}
\label{H_temperature}
\ee

and

\beq
J(\xi_{1}, \xi_{2}, x, y)&=& \pi \frac{\Gamma\left(\xi_{1}+\xi_{2}\right)}{\sinh{\left( \frac{x}{2 y}\right)}}\Im\left\{ \frac{e^{-2 \pi \xi_{1} \tau}}{\Gamma(\xi_{2})}\frac{e^{-ix}}{\Gamma\left( 2+i \frac{x}{2\pi y}\right) \Gamma\left( \frac{(\xi_{1}+\xi_{2})}{2}-i \frac{x}{2\pi y}\right)}{}_{2} F_{1} \left( \xi_{1}, \frac{(\xi_{1}+\xi_{2})}{2}+i \frac{x}{2\pi y}, 2+ i \frac{x}{2\pi y}; e^{-4 \pi y}\right)\right. \nonumber \\
&&\left.   -\frac{e^{-2 \pi \xi_{2} y}}{\Gamma(\xi_{1})}\frac{e^{i x}}{\Gamma\left(-i \frac{x}{2\pi y}\right) \Gamma\left(\frac{(\xi_{1}+\xi_{2})}{2}-i \frac{x}{2\pi y}\right)}{}_{2} F_{1} \left(\xi_{2}, \frac{(\xi_{1}+\xi_{2})}{2}-i \frac{x}{2\pi y}, - i \frac{x}{2\pi y}; e^{-4 \pi y}\right)\right\},
\label{J_temperature}
\eeq

\end{widetext}
where $\Im \left\{...\right\}$ indicates the imaginary part and ${}_{2} F_{1}\left(a, b, c ;z\right)$ is the hypergeometric function.\cite{Gradshteyn94}

The finite temperature electron Green's function is (for $ \omega, k_{B} T \ll  \omega_{c}$)

\beq
\tilde{\mathcal{P}}_{\xi}(\omega, T)&=& \left (\frac{2\pi k_{B}T}{\omega_c}\right )^{\xi-1}\frac{e^{-\frac{ \omega}{2 k_{B}T }}}{\omega_c} \nonumber \\
&& \mathcal{B}\left [\frac{\xi}{2}-i\frac{ \omega}{2\pi k_{B}T },\frac{\xi}{2}+i\frac{ \omega}{2\pi k_{B}T }\right ]
\eeq

being $\mathcal{B}\left[a, b\right]$ the Euler Beta function.\cite{Gradshteyn94}\\

\section*{Acknowledgements}

We thank A. Braggio and N. Magnoli for useful discussions. The support of CNR STM 2010 program, EU-FP7 via Grant No. ITN-2008-234970 NANOCTM and CNR-SPIN via Seed Project PGESE001 is acknowledged.


\begin{thebibliography}{10}
\bibitem{NoteH} Note that this Hamiltonian is formally analogous to the one of a  spinless Luttinger liquid. This is due to the fact that in this system the spin degree of freedom is redundant once identified the direction of propagation (right or left) along the edge.
\bibitem{note-prob} According to the Eqs. 12 in Ref.\onlinecite{Dolcini11}, the spin-preserving (-flipping) tunneling event occours with probability $\mathcal{P}_{sp}=4\gamma_{sp}^2/(1+\gamma_{sp}^2+\gamma_{sf}^2)^2$ ($\mathcal{P}_{sf}=4\gamma_{sf}^2/(1+\gamma_{sp}^2+\gamma_{sf}^2)^2$), where the dimensionless parameter $\gamma_{sp/sf}$ depends on the tunneling amplitude $\lambda_{sp/sf}$ according to the relation $\lambda_{sp/sf}=2v_F \gamma_{sp/sf}$. Hence the ratio between the two probabilities is $\mathcal{P}_{sp}/\mathcal{P}_{sf}=(\gamma_{sp}/\gamma_{sf})^2$ and depends on the local spin-orbit coupling.
\bibitem{note_universality} Note that, when the pumping frequency $\omega$ is much smaller than $v_F/2a$ (adiabatic limit), the two QPCs are equivalent to a single impurity positioned midway ($x=0$).
\bibitem{Note3b} Note that this value of $\omega_{0}$ guarantees that the relevant features of the pumping can be observed within the regime of validity of the helical Luttinger liquid picture.
\end{thebibliography}
\end{document}